\documentclass[useAMS,usenatbib]{mn2e}

\title[Nulls and mode-changes]{A bistable pulsar magnetosphere: nulls and mode-changes}
\author[P. B. Jones]{P. B. Jones\thanks{E-mail:
p.jones1@physics.ox.ac.uk}\\
University of Oxford, Department of Physics, Denys Wilkinson Building, 
Keble Road, Oxford OX1 3RH, England\\}
\begin{document}

\date{}

\pagerange{\pageref{}--\pageref{}} \pubyear{}

\maketitle

\label{firstpage}

\begin{abstract}
It is shown that the ion-proton magnetosphere is unstable in a limited area of the 
$P - \dot{P}$ plane against transitions to a self-sustaining inverse Compton scattering mode in which the particles accelerated are mainly protons with a small component of positrons.  It is argued that this mode cannot be absolutely stable.  The number density of any outward-moving pair plasma is small and electron and positron Lorentz factors too high to support growth of any collective mode capable of exciting normal pulsar coherent radio emission.  Particle fluxes and the position at which they pass through the light cylinder are mode-dependent and in principle, transitions can be accompanied by changes in spin-down torque.  The properties of the system are discussed in relation to observations of nulls, mode-changes, and the group of long-term intermittent pulsars.
\end{abstract}


\section{Introduction}

Radio pulsar observations in the last half century have revealed many common but not universal phenomena that are not as yet perceived as understood, of which nulls and mode-changes in the emission are examples.  In the same period, knowledge of magnetospheric structure has progressed little beyond the early insights of Goldreich \& Julian (1969), Radhakrishnan \& Cooke (1969) and Ruderman \& Sutherland (1975) except for recognition of the role of the Lense-Thirring effect and for
 the application of modern numerical plasma-physics techniques in the force-free approximation to the region near the light cylinder (see, for example, Bai \& Spitkovsky 2010 and Brambilla et al 2018 who also survey more recent work).

Our view is that the failure to make progress is a consequence of two factors.  Firstly, the proposition that neutron stars with negative polar-cap corotational charge density (${\bf \Omega}\cdot{\bf B} > 0$, where ${\bf \Omega}$ is the rotational spin and ${\bf B}$ the polar-cap magnetic flux density: Goldreich \& Julian 1969) can support the observed phenomena is, in Bayesian terms, one of zero prior probability.  Secondly, in the ${\bf \Omega}\cdot{\bf B} < 0$ case, failure to recognize that the generation of protons by the reverse flux of electrons at the polar cap is a dominant factor in determining the plasma composition.

In connection with our first point, electron motion above the polar cap is limited.  Electrons are in the Landau ground-state and drift velocities $-c({\bf E}\times{\bf B})/B^{2}$ are very small.  The electron work function at the neutron-star surface is so small that a space-charge-limited flow boundary condition is valid at all times, equivalent to a Dirichlet boundary condition on the surface and, in some approximation, on the surface separating open from closed sectors of the magnetosphere.  The system of accelerated electrons is described solely by the Vlasov-Maxwell equations except that pair creation is necessary for the formation of any collective mode that might lead to coherent radio-frequency emission.  It is difficult to understand how single-photon pair creation in the whole population of radio-loud pulsars, including the millisecond pulsars, can produce an electron-positron plasma of the required density or is even possible (see, for example, Hibschman \& Arons 2001, Harding \& Muslimov 2002).  The degrees of freedom in the ${\bf \Omega}\cdot{\bf B} > 0 $ case are so limited that we consider the assignment of zero prior probability justified.

The generation of protons in the positive corotational charge-density ${\bf \Omega}\cdot{\bf B} < 0$ case introduces new degrees of freedom to the system.  The origin of the acceleration field ${\bf E}_{\parallel}$ above the polar cap has been explained in terms of the Lense-Thirring effect in important papers by Beskin (1990) and Muslimov \& Tsygan (1992).  It is the interrelation between this field and its screening by photo-ionization of accelerated ions that makes the ${\bf \Omega}\cdot{\bf B} < 0$ system so complex but interesting.  It has been emphasized previously (Jones 2016; also papers cited therein) that these processes are the source of the coherent radio emission observed at frequencies of the order of $1$ GHz.

The introduction of new degrees of freedom brings with it additional parameters, specifically the blackbody surface temperature and the ionic atomic number, about which little is known.  Thus the magnetosphere is essentially a complex system in the sense described in a recent essay by Kivelson \& Kivelson (2018) and consideration should be given to what can realistically be expected from a model or theory of it.

Thus the work in this paper assumes that most, or possibly all, radio-loud pulsars, including those displaying nulls or mode-changes have ${\bf \Omega}\cdot{\bf B} < 0$.
We shall attempt to show that two distinct modes differing in average values of $E_{\parallel}$ and of ion, proton, and positron flux densities may exist quite naturally in limited regions of the pulsar $P-\dot{P}$ plane.  It is possible to see how each mode can be unstable against transition to the other.  Observations indicate that the timing of such transitions is essentially, though not always, chaotic, but this paper is limited to demonstrating the nature of the modes and the origin of their instability.

The radio-loud mode is believed, with some observational evidence (Jones 2016), to consist of an ion-proton plasma in which $E_{\parallel}$ is largely screened by the reverse flux of photo-ionization electrons.  The plasma must satisfy certain conditions if the Langmuir mode is to have a growth rate sufficient to produce strong plasma turbulence.  These are summarized in Section 2.

For convenience, the second mode can be referred to as radio-quiet, though observations show that this is not always strictly the case. It is argued here that this mode is most probably a self-sustaining inverse Compton scattering process generating a modest flux of electron-positron pairs.  This has not been considered previously in relation to the ${\bf \Omega}\cdot{\bf B} < 0$ case, and is described in detail in Section 3.

Contrary to what has been the canonical view, the presence of electron-positron pairs in this mode does not lead to coherent radio emission of the observed intensities in the $1$ GHz region, but may be significant in relation to incoherent emission, possibly at X-ray frequencies, for which a leptonic source is essential.

There are extremely large numbers of publications describing the observed properties of nulls and mode-changes, mostly in individual pulsars and we refer at this stage to a recent paper (Lyne et al 2017) which contains a useful and succinct summary.  But in Section 4 the properties of the second mode of Section 3 are discussed in relation to a number of individual pulsar observations and it is argued that they provide a basic physical framework for understanding the phenomena.

\section{The radio-loud state}

The properties of the ion-proton plasma mode have appeared in a number of previous publications (see Jones 2016) but are summarized here in order that they may be distinguished from the nominally radio-quiet state described in Section 3.  The basic requirement is that the longitudinal or quasi-longitudinal Langmuir mode should have an amplitude growth factor $\exp\Lambda$ sufficient to generate strong plasma turbulence with $\Lambda \approx 30$ adopted as a working value.  It is,
\begin{eqnarray}
\Lambda = \frac{Ra_{\Lambda i}}{c}\int^{\eta_{e}}_{1}\gamma^{-3/2}_{i}(\eta)
\omega_{pi}(\eta)d\eta
\end{eqnarray}
taken over particle paths to the emission region $\eta_{e}$.  Here $a_{\Lambda i} \approx 0.2$ is a dimensionless constant, $\gamma_{i}$ is the Lorentz factor and $\omega_{pi}$ is defined in terms of observer-frame variables for particle $i$,
\begin{eqnarray}
\omega^{2}_{pi} = \frac{4\pi n_{i}Z^{2}_{i}{\rm e}^{2}}{m_{i}},
\end{eqnarray}
with $m_{i}$ and $Z_{i}$ as the mass and charge of the ion, and $\eta$ the radius in units of the neutron-star radius $R$ (Jones 2012).  The conditions are as follows.

(i)		Plasma accelerated must contain both proton and ion components.

(ii)	The transverse width of the bundle of flux lines on which the plasma flows must be at least as large as the Langmuir-mode rest-frame wavelength.

(iii)	Screening of $E_{\parallel}$ within the bundle must be such as to limit the Lorentz factors to moderate, though relativistic, values ($\gamma \sim 20$).  This is facilitated by ions having high enough atomic number to remain partially ionized during their acceleration at $\eta < \eta_{e}$.  Thus a plasma of deuterons and protons in which there can be no screening would in general be too rapidly accelerated to have an adequate $\Lambda$.

(iv)	Owing to the small lepton mass, a positron component distributed with Lorentz factors overlapping those of the ions and protons can modify the dielectric tensor to the extent of extinguishing the Langmuir mode growth-rate (Jones 2015).

\section{Inverse Compton scattering}

Inverse Compton scattering by a Goldreich-Julian flux of primary electrons has been considered by many authors as a source of secondary pairs in an ${\bf \Omega}\cdot{\bf B} > 0$ magnetosphere (see, for example, Hibschman \& Arons 2001).  There is no easily quantifiable source of primary leptons in the ${\bf \Omega}\cdot{\bf B} < 0$ case: the most probable is the class of neutron capture reactions $n +(A,Z)\rightarrow (A+1,Z)+\gamma$ in which the neutrons originate in the same photo-nuclear reactions as the protons and at approximately the same rate.  The neutrons of $2-3$ MeV energy are not thermalized as are the protons but scatter with a cross-section of the order of $1$ bn so that a fraction of them will diffuse rapidly towards the surface.

Capture in this region produces $\gamma$-rays which can enter the open magnetosphere above the polar cap.  Multiple Compton scatters within the shower itself are a further possible source of such outward-directed photons.  But this also is difficult to quantify. The single-photon attenuation coefficient through pair production is,
\begin{eqnarray}
\frac{m{\rm e}^{2}B(\eta)\sin\theta}{2\hbar^{2}B_{c}}\left(0.377\exp\frac{-4}{3\chi}\right)
\end{eqnarray}
(Erber 1966) in which, $\chi = k_{\perp}B(\eta)/(2mcB_{c})$, where $k_{\perp}$ is the photon momentum component perpendicular to ${\bf B}$ and $B_{c} = m^{2}c^{3}/{\rm e}\hbar =4.41\times 10^{13}$ G.  The angle between ${\bf k}$ and ${\bf B}$ is $\theta$. A value $\chi = 0.070$, equivalent to $k_{\perp}B_{12} = 6mc$ corresponds with attenuation of $10^{-2}$ cm$^{-1}$.  A relatively narrow window of values in the vicinity of this would give attenuation neither so small that the $\gamma$ escapes from the open region of the magnetosphere nor so large that the positron formed annihilates in the  neutron-star atmosphere.  Polar cap magnetic flux densities of $2-3\times 10^{12}$ G would satisfy this condition for typical nuclear-capture $\gamma$-ray energies.  Consequently the source would not function in millisecond pulsars or in high-field neutron stars.  Whilst the rate of positron formation is contingent on many factors which are not well known, it is approximately a linear function of reverse-electron energy input to the polar cap:  it is denoted here by $W_{e}$, the number of positrons formed above the neutron-star atmosphere within the open magnetosphere per unit reverse-electron energy.

For an outward accelerated positron at $\eta$ with Lorentz factor $\gamma$ and velocity $\beta$, the transition rate from a blackbody radiation field of temperature $T_{s}$ is,
\begin{eqnarray}
\Gamma = 2\pi\int^{\theta_{max}}_{0}d(\cos\theta)\int^{\infty}_{0}d\omega \hspace{2cm} \\   \nonumber
\hspace{2cm}  \left(
\frac{\omega^{2}\tilde{n}(\omega, T_{s})}{4\pi^{3}\hbar^{3}c^{2}}\right)
\left(1 - \beta\cos\theta \right)\sigma(s),
\end{eqnarray}
in which $\cos \theta_{max} = (\eta^{2} - 1)^{1/2}/\eta$, the photon occupation number is $\tilde{n}$ and $\sigma(s)$ is a partial Klein-Nishina cross-section at Lorentz-invariant total energy-squared $s$,
\begin{eqnarray}
s = m^{2} + 2m\gamma\omega(1 - \cos\theta),
\end{eqnarray}
and $\omega$ is the blackbody photon energy.  In the Klein-Nishina region of $s \gg m^{2}$, the cross-section is dominated by backward scattering owing to the u-channel pole and we therefore adopt the approximate expression for scattering into the backward hemisphere in the centre-of-momentum system,
\begin{eqnarray}
\sigma(s) = \pi\left(\frac{{\rm e}^{2}}{mc^{2}}\right)^{2}\left(\frac{2m^{2}}{s}
\ln\frac{s}{m^{2}} - \frac{m^{2}}{s}\right)
\end{eqnarray}
for $s \geq 5$, with linear interpolation for $1 \leq s \leq 5$.  For the same  reason, and following Hibschman \& Arons (2001), we assume the scattered photon energy to be the maximum,
\begin{eqnarray}
ck_{1} = m\gamma\left(1 - \frac{m^{2}}{s}\right)   \hspace{2cm} \gamma \gg 1
\end{eqnarray}

The direction of ${\bf k_{1}}$ is initially closely tangential to the local ${\bf B}$ and it is the increase in this angle as the photon propagates that determines the point at which pair creation occurs.  The contribution of aberration to this angle is an order of magnitude smaller than that of the flux-line curvature and we neglect it.  The intersection of a photon emitted tangentially at $\eta$ with a flux line at $\eta^{\prime}$ is at  an angle,
\begin{eqnarray}
\frac{3u(\eta)}{4R\eta}\left(1 - \frac{\eta}{\eta^{\prime}}\right),
\end{eqnarray}
for dipole field geometry.   Photon conversion to first-generation pairs occurs in the interval $\eta < \eta^{\prime} \leq 4\eta/3$ provided its momentum satisfies,
\begin{eqnarray}
k_{1} > k_{c} = \left(\frac{6mc}{B_{12}(1)}\right)\left(\frac{4\eta}{3}\right)^{3}
\left(\frac{16R\eta}{3u(\eta)}\right),
\end{eqnarray}
in which $u$ is the lateral displacement of the Compton event from the magnetic axis. 
The electron and positron are in high Landau states radiating by synchrotron emission. We adopt the Hibschman \& Arons estimate of $0.22k_{1}/k_{c}$ for the number of higher generation pairs thereby produced.

The unscreened acceleration potential is assumed to be,
\begin{eqnarray}
  \hspace{1cm} V_{max}(\eta, u) =  \hspace{4cm} \nonumber   \\ 1.25\times 10^{3}\left(1 - \frac{1}{\eta^{3}}\right)
\left(1 - \frac{u^{2}}{u^{2}_{0}}\right)\frac{B_{12}(1)}{P^{2}} \hspace{4mm} {\rm GeV}
\end{eqnarray}
(see Jones 2013) in which $u_{0}$ is the radius of the circular open magnetosphere at $\eta$, and $P$ the rotation period.

Table 1 gives the mean reverse-electron energy input from a single positron in a partially-screened acceleration field $s_{V}V_{max}$.  Inputs $\epsilon_{1}$ and $\epsilon^{ICS}$ from the first and from all generations of electrons are shown separately for values of $u = s_{u}u_{0}$.  A distinction from Hibschman \& Arons is that we neglect screening of the Lense-Thirring potential by ICS reverse electrons. This is justified by the fact that for ${\bf \Omega}\cdot{\bf B} < 0$, the proton and positron current densities satisfy $J^{p} \gg J^{e}$: typically $J^{e}$ is of the order of $10^{-1}J^{p}$ and the reverse electrons from photo-ionization are the principal source of screening.  The approximations on which Table 1 is based are such that the values of $\epsilon^{ICS}$ can be no more than a guide.  The most serious, the assumption of dipole-field geometry, is unavoidable. 

\begin{table}
\caption{This shows separately in units of GeV, the reverse-electron energy inputs per positron accelerated for first $(\epsilon_{1})$ and all generations of pairs $\epsilon^{ICS}$ as functions of the scale parameter $s_{V}$ which represents the degree of screening of the acceleration potential so that $V = s_{V}V_{max}$, $V_{max}$ being defined by equation (10). Columns 2 to 7 give these energies for positrons accelerated at radii $u = s_{u}u_{0}$ from the magnetic axis in a circular polar cap of radius $u_{0}$, with $s_{u} = 0.2, 0.5, 0.8$.  The neutron-star parameters are $P = 0.81$ s and $B_{12} = 2.6$ for PSR 1931+24.  The temperatures are $T_{s} = 2.5, 5.0, 10.0\times 10^{5}$ K in ascending order from the foot of the Table.}

\begin{tabular}{@{}lrrrrrr@{}}
\hline
$V$ & $\epsilon_{1}$ & $\epsilon^{ICS}$ & $\epsilon_{1}$ & $\epsilon^{ICS}$ & $\epsilon _{1}$ & $\epsilon^{ICS}$  \\
\hline
$V_{max}$   & GeV  & GeV  & GeV  & GeV  & GeV  & GeV   \\
\hline
   &  $s_{u}$ = 0.2  &  0.2  &  0.5  &  0.5  &  0.8  &  0.8     \\
\hline
1.00  &  17.3  &  404.1 & 16.3 & 731.9 & 13.5 & 472.1  \\
0.50  &  14.6 & 179.4 & 13.6 & 315.5 & 10.9 & 197.8  \\
0.20  &  11.2 & 62.1 & 10.2 & 101.9 & 7.6 & 60.9  \\
0.10  &  8.7 & 28.7 & 7.8 & 43.1 & 5.4 & 24.6  \\
0.05  &  6.5  & 13.9 & 5.6 & 18.3 & 3.5 & 9.9  \\
0.02  &  3.9 & 5.6 & 3.2 & 6.1 & 1.7 & 2.9  \\
0.01  &  2.0 & 2.5 & 1.8 & 2.6 & 0.7 & 1.0  \\

\hline
1.00  &  3.6   & 86.0   & 3.4   & 154.4   & 2.7   & 96.2        \\
0.50  &  3.0  &  37.1  &  2.8  &  64.4  &  2.1  &  38.6     \\
0.20  &  2.2  &  12.1  &  1.9 & 19.6  &  1.3  &  10.9     \\ 
0.10  &  1.6  &  5.3  &  1.4  &  7.8  &  0.9&  4.0     \\
0.05  &  1.1 & 2.4  &  0.9  &  3.0  &  0.5  &  1.4    \\
0.02  &  0.6  &  0.8  &  0.4  &  0.8  &  0.2  & 0.3   \\
0.01  &  0.3  &  0.3  &  0.2  &  0.3  &  0.0  &  0.1   \\

\hline
1.00  &  0.7 & 17.8 & 0.7 & 31.5 & 0.5 & 18.8  \\
0.50  &  0.6 & 7.4 & 0.5 & 12.6 & 0.4 & 7.1  \\
0.20  &  0.4 & 2.2 & 0.3 & 3.5 & 0.2 & 1.8  \\
0.10  &  0.3 & 0.9 & 0.2 & 1.3 & 0.1 & 0.6   \\
0.05  &  0.2 & 0.4 & 0.1 & 0.4 & 0.1 & 0.2  \\
0.02  &  0.1 & 0.1 & 0.0 & 0.1 & 0.0 & 0.0  \\
0.01  &  0.0 & 0.0 & 0.0 & 0.0 & 0.0 & 0.0  \\
\hline  
\end{tabular}
\end{table}

Following Harding \& Muslimov (2002), we adopt an open magnetosphere radius,
\begin{eqnarray}
u_{0}(\eta) = \left(\frac{2\pi R^{3}\eta^{3}}{cPf(1)}\right)^{1/2}
\end{eqnarray}
where $f(1) = 1.368$ for a neutron star of $1.4$ $M_{\odot}$ and radius $R = 1.2\times 10^{6}$ cm.  The surface temperature $T_{s}$ is of that part of the surface visible from a radius $\eta$.  The temperature dependence of $\epsilon^{ICS}$ is largely determined by the properties of blackbody radiation and we find $\epsilon^{ICS} \propto T_{s}^{2.3}$ for values of $s_{V} \sim 0.5$ that are relevant to self-sustaining ICS.
The rotation period $P = 0.81$ s and polar-cap $B_{12}(1) = 2.6$ have been chosen to coincide with those of PSR 1931+24  (Manchester et al 2005).

The production rate for protons is $W_{p}\epsilon^{ICS}$ per positron.  Values of $W_{p}$ are approximately proportional to $Z^{-1}$, where $Z$ is here the mean atomic number of nuclei undergoing photo-absorption in the electromagnetic showers. Typical values are $0.2 - 0.5$ GeV$^{-1}$ (see Jones 2010, Table 1) and for values $s_{V} \approx 0.5$, of the order of $10$ protons are produced per positron at $5\times 10^{5}$ K.

The particle flux can consist of ions, positrons and protons: but in what order of priority do they comprise a flux at the Goldreich-Julian value?  Positrons are formed at low altitudes ($< u_{0}$) above the polar cap and thus are bound to be part of the flux.  Protons are not in static equilibrium in the predominantly ion atmosphere in local thermodynamic equilibrium (LTE) and thus having a higher charge-to-mass ratio enter the flux in preference to ions.  An excess of protons has time to fractionate forming a further LTE layer at the top of the atmosphere.  Ions enter the
flux and undergo photo-ionization to a charge $Z_{\infty}$ only if the protons are insufficient to form such a layer or produce a Goldreich-Julian charge density.  It is virtually impossible to estimate with confidence the value of $W_{e}$, but we adopt a conservative value $W_{e} = 10^{-1}W_{p}$.  In this case it is clear from the $\epsilon^{ICS}$ in the Table that self-sustaining ICS pair creation would be possible given favourable conditions.  These are: a polar-cap $B$ within the narrow window suitable for a neutron-capture $\gamma$-ray attenuation rate of $10^{-1} - 10^{-4}$ cm$^{-1}$, a sufficient $T_{s}$ and $V_{max}$, as may be the case in PSR 1931+24.  The high values of $\epsilon^{ICS}$ are a consequence of higher generations of pairs created in regions of $V \rightarrow V_{max}$.  But for $0.01 < s_{V} \leq 0.1$, typical of the mean potential in the radio-loud mode, it is clear that $\epsilon^{ICS}$ is not significant in comparison with the energy input from photo-ionized reverse-electrons.

The description of processes at the polar cap can be formalized with the definitions
$\epsilon W_{p} = K$, where $\epsilon$ is the reverse-electron energy per unit charge of ion accelerated, and $\epsilon^{ICS}W_{p} = K^{ICS}$.  Then the ion, proton, and positron current densities $J^{z,p,e}$ are related by,
\begin{eqnarray}
J^{p}(t) + \tilde{J}^{p}(t) =   \hspace{4cm} \nonumber   \\\int^{t}_{-\infty}dt^{\prime}f_{p}(t - t^{\prime})
\left(KJ^{z}(t^{\prime}) + K^{ICS}J^{e}(t^{\prime}) \right)
\end{eqnarray}
in which $f_{p}$ is the proton diffusion function, normalized to unity and of scale given by a diffusion time $\tau_{p} \sim 1$ s.  The rate at which protons enter and add to any LTE layer that might exist is $\tilde{J}^{p}$.  Both $K$ and $K^{ICS}$ are functions of $V({\bf u},t^{\prime})$.  A similar type of relation describes positron formation, $\epsilon^{ICS}W_{e} = K^{e}$, but the time delays are of the order of particle transit times and are negligible compared with $\tau_{p}$.  Thus the expression reduces to,
\begin{eqnarray}
J^{e}(t)\left(1 - \frac{W_{e}}{W_{p}}K^{ICS}\right) = \frac{W_{e}}{W_{P}}KJ^{z}(t)
\end{eqnarray}
at any instant.  Equation (12) holds locally at any polar-cap coordinate ${\bf u}$ in respect of $J^{p}$, $\tilde{J}^{p}$, and $J^{z}$, and approximately so for $J^{p}$, $\tilde{J}^{p}$, and $J^{e}$, whilst equation (13), based on single-photon pair creation can be valid only for averages over the whole polar cap: equation (13)  fails and self-sustaining positron creation occurs at the threshold $W_{p} = W_{e}K^{ICS}$.

Equation (12) has been modelled in a very basic manner (Jones 2013) by representing the polar cap in terms of finite elements $\delta {\bf u}$ each contributing to the potential $V({\bf u},t)$, but with neglect of ICS.  It provides a simulation of a radio-loud polar cap, downward fluctuations in $V$ corresponding with radio emission in accordance with equation (1) and condition (ii).  As a mode it is stable under conditions in which $\tilde{J}^{p} = 0$.

But large upward fluctuations of $V$ from its time average occur in the model at instants when $J^{z} = 0$ in all elements and $V \rightarrow V_{max}$.  These intervals which are nulls are usually brief because $J^{z} = 0$ in $\delta {\bf u}$ means zero proton production in that element so that eventually its $J^{p}$ falls below $J_{GJ}$, the Goldreich-Julian  current density.  Thus ion emission re-commences and screening by the reverse electrons reduces $V$.  There is no transition to the ICS mode unless the threshold $W_{p} = W_{e}K^{ICS}$ is reached.  This depends on suitable values of a number of parameters (surface temperature $T_{s}$, $B$ and $V_{max}$) which is likely only in limited areas of the population $P - \dot{P}$ plane.

Upward fluctuations of $V$ extend over increasing areas of the polar cap if $V_{max}$ is large enough.  But provided there is a boundary condition on the cylindrical surface separating open from closed magnetospheres there must be an annular region with outer boundary $u_{0}$ in which $V$ is not large enough to produce a proton current density $J^{p} = J_{GJ}$.  Thus ion emission forms a part of the current density in this region but conditions (ii) and (iii) of Section 2 may not be satisfied, resulting in no observable radio emission.

If the mode persists with $\tilde{J}^{p} > 0$ in the central region of the polar cap, an LTE atmosphere of protons grows, possibly to the extent of developing a liquid phase.  Ion number densities at the surface of a neutron star have been calculated by Medin \& Lai (2006) and can be expressed as,
\begin{eqnarray}
N = 2.6\times 10^{26}Z^{-0.7}B^{1.2}_{12}   \hspace{1cm} {\rm cm}^{-3},
\end{eqnarray}
where $Z \geq 6$ is here the nuclear charge.
The radiation length defined in terms of the Bethe-Heitler formulae (Bethe 1934) for bremsstrahlung and pair production cross-sections, with modified screening appropriate for the magnetic field, is then
\begin{eqnarray}
l_{r} = 1.6 Z^{-1.3}B^{-1.2}_{12}\left(\ln(12Z^{1/2}B^{-1/2}_{12})\right)^{-1}
{\hspace{1cm}} {\rm cm},
\end{eqnarray}in which the bracketed term, replaces the $\ln(183Z^{-1/3})$ term of the Bethe-Heitler expression (see Jones 2010).  For protons, it may be preferable to use the linear-chain spacing found by Medin \& Lai, extended to a simple cubic lattice, giving a density
$N = 5.5\times 10^{26}B^{1.2}_{12}$, about a factor of two larger than equation (14) and therefore a radiation length $l^{p}_{r}$ smaller by the same factor.

The depth of any liquid phase is naturally limited to that necessary to contain the processes of the electromagnetic shower.  Also, the nature of the shower is modified because in a proton liquid, Compton scattering is the dominant reaction: the mean free path for this is small compared with $l^{p}_{r}$.  Furthermore, $W_{e}\rightarrow 0$ because there is no source of neutrons.

Even if the proton LTE atmosphere or liquid phase exists over a large fraction of the polar cap, it can have no long-term stability.  The reason is that within an element of area $\delta {\bf u}$, an ion current component screens effectively owing to the reverse-electron flux: protons or completely stripped ions of low $Z$ do not screen.
Thus in an annular region of ${\bf u}$ near $u_{0}$, $\tilde{J}^{p} = 0$ and equation (12) shows that a downward fluctuation in $K$ and $K^{ICS}$ at $t^{\prime}$ leads to a downward fluctuation in $J^{p}(t)$ and an increase in $J^{z}(t)$ which further increases screening in that region.  The system is one of extraordinary complexity and, as indicated in Section 1, this paper is limited to defining the conditions under which the two modes exist and showing how the transitions between them can occur.

Neutrons and protons are generated principally at the shower maximum at a depth of about $10$ radiation lengths.  The nuclear charge of ions entering the LTE atmosphere is therefore reduced to $Z_{s}$ from the assumed original $Z_{0} = 26$.  It has been shown previously (Jones 2011) that instability, in the form of a time-dependent $Z_{s}$ ensues if at any instant $Z_{s}$ is such that the ion is completely stripped whilst in the LTE atmosphere. In this case, there is no reverse-electron flux at $\delta {\bf u}$ so that eventually higher values of nuclear charge reach the surface and the reverse-electron flux re-commences.  the time-scale for this is given broadly by the elapsed time in which one radiation length of ions leave the surface in a Goldreich-Julian flux.  It is,
\begin{eqnarray}
t_{rl} = 2.1\times 10^{5}\left(\frac{P}{Z_{s}B_{12}\ln(12Z_{s}^{1/2}B^{-1/2}_{12}}\right)
{\hspace{5mm}}  {\rm s}.
\end{eqnarray} 
Chaotic behaviour of this nature is likely to contribute to transitions from ICS to the normal radio-loud state.

\section{Conclusions}

Application of the results found in Section 3 to specific pulsar observations is on the basis that an ion-proton plasma is the source of the coherent radio emission normally observed at frequencies of the order of $1$ GHz.  Electron-positron pairs in the ICS mode cannot lead to radio emission of the intensity usually seen principally because the lepton Lorentz factors are too large to allow a significant growth rate in any collective mode that could be the source.  Equation (9) gives $k_{c} \sim 10^{4}$ for the minimum momentum of a photon so that the lepton Lorentz factors, substituted into equation (1) would not give an adequate growth factor exponent, of the order of $\Lambda \approx 30$. Thus neither a complete ICS mode nor an area of polar cap with a short-term $\tilde{J}^{p} \neq 0$ can be a source of normal radio emission although we have not studied the possibility that some form of low-intensity emission might result.

Switching to or from a radio-loud state is known to occur within a rotation period. Fluctuations in the screening of $V_{max}$ are inherent in the model owing to reverse-electron fluxes and consequent proton production.  The basic time-scale here is the proton diffusion time from shower maximum to the surface, estimated to be of the order of $\tau_{p} \sim 1$ s, as in equation (12).  Proton fluxes provide no screening so that the time-scale for large fluctuations away from $V_{max}$ must be no greater than $\tau_{p}$.  Radio emission needs modest values of the local ion Lorentz factor, of the order of $10 - 30$, and hence large downward fluctuations from $V_{max}(\eta,u)$.
The dependence of the amplitude growth exponent $\Lambda$ on Lorentz factor in equation (1) shows that the maximum mode amplitude attained is an extremely sensitive function of acceleration potential a small change in which can result in failure to reach a turbulent state and the cessation of radio emission, or the reverse.  The rotation period of mode-changing pulsars is usually of the order of $P = 1$ s, hence the model is at least qualitatively consistent with observation.

Fluctuations downward from $V_{max}$ are a feature of the basic model polar cap studied by Jones (2013) and their magnitude is an increasing function of $T_{s}$.  Photo-ionization cross-sections are large and transitions occur promptly once blackbody photons in the rest-frame of an ion reach the threshold energy.  Thus the ion Lorentz factor at this point is $\propto  T_{s}^{-1}$.  Consequently, photo-ionization screens the Lense-Thirring potential very efficiently in young ${\bf \Omega}\cdot{\bf B} < 0$ neutron stars so that upward fluctuations necessary for a transition to the ICS mode have negligible probability.

The temperature $T_{s}$ is that of the surface visible from radius $\eta$ on the magnetic axis and is in the local proper frame.  Owing to this and to the ${\bf B}$-dependent anisotropy of thermal conductivity in the neutron-star crust, it will be larger than the whole-surface average seen by an observer.  Given these factors and on the basis of temperatures listed by \"{O}zel (2013) for a group of pulsars $\tau_{c} <1$ yr in age, we shall assume that $T_{s} = 5\times 10^{5}$ K is a reasonable estimate for any set of pulsars that have ages of the order of $1$ Myr including those showing long-term intermittency.
Here we refer to PSR 1931+24 (Kramer et al 2006), J1832+0029 (Lorimer et al 2012),
J1841-0500 (Camilo et al 2012), J1910+0157 and J1929+1357 (Lyne et al 2017).
Fluctuations in $V$ are larger at this temperature whilst the establishment of a long-term ICS mode remains possible.

Table 1 indicates that this ceases to be so at lower temperatures, which is consistent with the position in the $P- \dot{P}$ plane at which intermittency is seen.  At low $T_{s}$, where ICS modes cannot be self-sustaining, because photo-ionization cross-sections are large near thresholds an ion-proton plasma remains possible.  But higher Lorentz factors are needed to reach the thresholds: thus the time-averaged potential $\langle V({\bf u},t)\rangle$ is increased as in consequence are values of $K$.  On average, elements of polar-cap area supporting an ion-proton flux become more sparse (see the basic model in Jones 2013).  Emission is within a cone whose semi-angle with respect to the local source-region magnetic flux lines is finite and determined by the ion Lorentz factors.  Hence an observer sees emission from flux lines in a band across the polar cap and will register a null if there is no ion-proton current density within the band that satisfies conditions (i) - (iv) of Section 2.  This accounts for the increase in short ($< 10P$) nulls as a function of age (Wang et al 2007).

Longer nulls can be caused by the variability in the surface nuclear charge $Z_{s}$ on time-scales derived from equation (16) and described at the end of Section 3.   Here, if $\epsilon^{ICS}$ is negligible, and averaged over time at any ${\bf u}$ within the polar cap,
\begin{eqnarray}
Z_{0} - \langle Z_{s}\rangle = \langle KZ_{s}\rangle,
\end{eqnarray}
so that increasing $K$ is certainly accompanied by decreasing $\langle Z_{s}\rangle$ creating the conditions for instability described previously in Jones (2011).  In this stage of evolution, the mean value of $V$ increases until condition (iii) ceases to be satisfied and the Langmuir mode growth-rate is too small to produce strong turbulence.  It is the most likely reason for the cessation of emission at an age which is a function of the parameters $P$, $B_{12}$, $T_{s}$ and $Z_{0}$.  This is consistent with the observed density of pulsars in the logarithmic $P - \dot{P}$ diagram which indicates not a death-line but the fact that pulsar deaths occur from a  relatively early time, of the order of $10$ Myr onwards.

The position of the Rotating Radio Transients (RRAT) in the sequence is enigmatic.  For one half of those listed by Keane et al (2011) with values of $P$ and $B$, equation (10) gives $V_{max}(\infty,0) < 10^{3}$ GeV, which is likely to be too small to support a self-sustaining ICS mode.  Also, some values of $B_{12}$ are too large to facilitate the optimum capture $\gamma$-ray attenuation rates of $10^{-1}- 10^{-4}$ cm$^{-1}$.  It is not even obvious that the radio emission conforms with the normal ion-proton plasma expectation.  Karastergiou et al (2009)  were able to study a limited number of single pulses from J1819-1458 but found circular polarization that is too small to provide evidence for an ion-proton plasma source.  Time intervals of
$10^{2-4}$ s between successive intervals of emission  suggest involvement of the surface nuclear charge with time-scales given by equation (16).  However, in the case of very high fields ($>B_{c}$) low energy electron-positron pairs may be present and the Langmuir-mode dispersion relation so modified that growth is prevented.  We refer to Jones (2015) for details of this additional explanation for nulls.

Mode-changes are much less frequently observed than nulls (see the review of Wang et al 2007) and are more difficult to characterize.  But we can consider two separate classes: those involving a change in mean pulse profile only and a very small number of cases in which X-ray emission is also observed.  The former are the more numerous and have been reported most recently by Lyne et al (2010; 6 pulsars), Young et al (2014; J1107-5907), Sobey et al (2015; B0823+26), and Brook et al (2016; 9 pulsars).   With the exception of J1107-5907, these are all less than or of the order of $1$ Myr in age, and 4 are listed in the Second Fermi LAT Catalogue of $\gamma$-emitting pulsars (Abdo et al 2013).

Lyne et al associated profile changes directly with changes in spin-down rate $\delta\dot{P}$ measured over intervals of $100 - 400$ d.  The later paper of Brook et al is of particular interest because it looked for changes in the relative flux density of different components in the profile of each pulsar.  These authors found profile variability but, except in the case of J1602-5100 with $\delta\dot{P}/\dot{P} = 0.05$ over a $600$ d interval, were unconvinced that a simple two-state spin-down model best fitted the data.  Values of $B_{12}P^{-2}$ for each of these pulsars with a temperature $T_{s} = 5\times10^{5}$ K are adequate for the formation of an ICS state, as described in Section 3. (The only exception is J1107-5907 with $B = 4.8\times 10^{10}$ G which is at least an order of magnitude too small to support the pair creation process described there.)  Formation of this state changes both the rate and the position at which particles pass through the light cylinder.  The net charge on the star must remain approximately constant on time-scales of the order of $P$.  This can be achieved by particle fluxes in the open sector of the magnetosphere including the outer gap, possibly also by a return current.  It is likely that an ICS state forms over an area of the polar cap with values of $u$ near the maximum in $\epsilon^{ICS}$, referred to as a partial ICS mode, with some ICS pair creation.
This alters the flux and nature of particles reaching the light cylinder in the beam accelerated from the polar cap.  Simultaneously, the flux of particles leaving through acceleration in the outer gap must change in order to maintain the overall charge balance.  The very high $\gamma$-ray luminosity of the older neutron stars listed by Abdo et al (2013) shows that particle acceleration in the outer gap can be the origin of a large fraction of the spin-down torque.  Thus changes caused by  a transition to even a partial ICS mode appear capable causing the small $\delta\dot{P}/\dot{P}$ observed.

Mode changes in pulsars with observed X-ray emission are very limited in number.  Hermsen et al (2017) have observed radio and X-ray emission simultaneously in PSR B1822-09.  Mode changes are seen in the radio with a mean separation of about $200$ s but no mode changes, synchronous or otherwise, appear in the X-rays.  The outstanding \textit{sui generis} pulsar is B0943+10.  The existence of synchronous mode-changes in radio and X-ray emission was first observed by Hermsen et al (2013) and then further by Mereghetti et al (2016). Of the two modes, the B-mode has the stronger radio flux and the weaker X-ray flux, and the Q-mode vice-versa.  Emission of X-rays is incoherent and requires a leptonic source.  The period is $P = 1.10$ s and $B_{12} = 2.0$, with $\tau_{c} = 5.0$ Myr:  ICS formation is possible given the criteria adopted here and it would be tempting in view of ICS electron-positron production to assign the Q-mode to a partial ICS state.  But that would introduce the problem of understanding the origin of the Q-mode radio emission which, although of approximately half the B-mode intensity, has a very different single-pulse structure consisting of sporadic bright pulses distributed with chaotic phases within the integrated profile (Bilous et al 2014).  It remains an enigmatic object.

Mode changes as described here must certainly change particle fluxes at the light-cylinder radius and its relation with recent work (Brambilla et al 2018 and work cited therein) would potentially be of interest.  Unfortunately, these plasma-physics computational techniques appear to have been applied exclusively to the ${\bf \Omega}\cdot{\bf B} > 0$ case with various conditions on particle injection so that a useful comparison between inner and outer magnetosphere work is difficult. 

The intermittent pulsars are cases in which ICS has an unambiguous role.  Values of $P$ and $B_{12}$ can support a self-sustaining ICS state according to the criteria of this paper, as in Table 1 which refers to B1931+24.  All members of the set are very closely positioned in the logarithmic $P - \dot{P}$ plane. Time-scales can be as long as $10^{7-8}$ s, several orders of magnitude larger than $\tau_{rl}$, and difficult to associate with any other obvious aspect of neutron-star physics.

The ICS mode has been demonstrated here as a state of the magnetosphere with the possibility of long-term but not absolute stability.  The approximate halving of  spin-down rate correlated with the transitions (Lyne et al 2006, 2017) requires a significant change in magnetospheric structure in the vicinity of the light cylinder to alter the torque.  The ICS mode certainly makes such changes possible because as we noted previously, the flux of particles and the position at which they pass through the light cylinder are substantially modified.  But further consideration of this is far beyond the scope of this paper.  More generally, it must be admitted that some objects remain enigmatic in the light of the ICS mode study described here.  The present work does rely on parameters which certainly exist but are not well known.  Nonetheless, it provides a physical understanding of nulls and mode-changes based on the nature of the plasma as discussed in the early paragraphs of Section 1.

\section*{Acknowledgments}

It is a pleasure to thank the anonymous referee for questions that have much improved the presentation of this work.

\bsp

\label{lastpage}

\end{document}